\documentstyle[preprint,aps,prd,epsf]{revtex}
\begin{document}
\newcommand{\eq}{\begin{eqnarray}}
\newcommand{\en}{\end{eqnarray}}
\newcommand{\shw}{(\omega^2-1)^{1/2}}
\baselineskip 10pt

\draft 
\title{\hfill MZ-TH/98-13\\
Charm and Bottom Baryon Decays in the
Bethe-Salpeter Approach: Heavy to Heavy Semileptonic Transitions}
%\vspace*{-0.2cm}
\author{M.\ A.\ Ivanov}
\address{Bogoliubov Laboratory of Theoretical Physics, Joint Institute
for Nuclear Research, 141980 Dubna, Russia}
%\vspace*{-0.2cm}
\author{J.\ G.\ K\"{o}rner}
\address{Johannes Gutenberg-Universit\"{a}t, Institut f\"{u}r Physik,
D-55099 Mainz, Germany}
%\vspace*{-0.2cm}
\author{V.\ E.\ Lyubovitskij}
\address{Bogoliubov Laboratory of Theoretical Physics, Joint Institute
for Nuclear Research, 141980 Dubna, Russia
and Department of Physics, Tomsk State University,
634050 Tomsk, Russia}
%\vspace*{-0.2cm}
\author{A.\ G.\ Rusetsky}
\address{Bogoliubov Laboratory of Theoretical Physics, Joint Institute
for Nuclear Research, 141980 Dubna, Russia
and HEPI, Tbilisi State University, 380086 Tbilisi, Georgia}
%\date{\today}

\maketitle
\vspace*{-1.cm}
\begin{abstract}\widetext
Charm and bottom baryons and mesons are studied within the framework of a
relativistically covariant 3D reduction of the Bethe-Salpeter equation.
We carry out an analysis of semileptonic decays of heavy hadrons within this 
framework using explicit oscillator-type wave functions where we calculate 
Isgur-Wise functions, decay rates and asymmetry parameters. Within this model 
we also study the effect of interactions between the light quarks inside the 
heavy baryon and how they affect the values of the computed heavy baryon 
observables. We also elaborate on the role of relativistic effects in the 
calculation of the heavy baryon Isgur-Wise function. 
\end{abstract}

\pacs{PACS number(s): 11.10.St, 12.39.Ki, 13.30.Ce, 14.20.Lq, 14.20.Mr}

\baselineskip 23pt

%\narrowtext
\section{Introduction}

There has been remarkable progress in the experimental study of hadrons
containing a single heavy quark \cite{PDG}. The experimental progress calls
for the development of theoretical approaches that allow one to study bound
systems of a heavy quark and light quarks/antiquarks. This would enable one
to analyse different weak decay processes (leptonic, semileptonic and
nonleptonic) of heavy baryons and heavy mesons on an equal footing. All
information about heavy hadron decays is contained in a set of reduced
form factors which are governed by the dynamics of their light constituents.
Since the momentum dependence of these reduced form factors cannot yet be
determined from first principles in QCD one has to turn to QCD-inspired
model studies of these quantities. Such models should take into account the 
full content of the symmetries of the underlying strong interaction Lagrangian
as e.g. the leading order spin-flavor symmetry of the HQET Lagrangian
\cite{Neubert}.

Any model devised for the quantitative description of heavy hadron weak
transitions should include relativistic effects. First, the average momenta
of light quarks inside the hadron are of the order of the constituent quark
mass. As a result, there are large relativistic effects in the dynamics of
heavy meson decays (see, e.g. \cite{Close,LeYaouanc}). Moreover, a general
model approach should also be applicable to the description of those decays of
heavy hadrons which are accompanied by a large momentum (energy) transfer.
Thus, one needs a genuine relativistic treatment of the problem under study.
There exist various relativistic approaches which enable one to study heavy
hadron weak transitions: QCD sum rules \cite{Neubert,Grozin,Dai_SR},
QCD on the lattice \cite{Lattice_1,Lattice_2,Lattice_3}, relativistic quark
models
\cite{DESY,Ivanov,Kroll,Konig,Mainz1,Mainz2,Faustov,Galkin,MesonIW,Huang,Guo,Cheng_cone,Roberts_Dubna,Roberts,Kalinovsky,Reinhardt,Mitra1,Mitra2,Mitra3,Mizutani}
including the approaches based on the use of the Bethe-Salpeter (BS) equations
\cite{Faustov,Galkin,MesonIW,Huang,Roberts_Dubna,Roberts,Kalinovsky,Mitra1,Mitra2,Mitra3}.

The BS formalism provides a systematic field-theoretical basis for the
treatment of bound-state to bound-state transitions in which the
interaction operators between the constituents can, in principle, be
constructed from the underlying Lagrangian of the theory. The presence of
confining interactions in the bound system precludes a straightforward use of
perturbation theory for the calculation of the BS kernel. One must necessarily
make a nonperturbative model ansatz for the kernel. In particular, utilizing
the skeleton expansion of the interquark kernel together with a plausible
approximation for the long-range behavior of the gluon propagator responsible
for the confinement of quarks one can obtain a coupled set of Schwinger-Dyson 
and Bethe-Salpeter equations in Euclidean space. This approach 
can be solved to compute hadronic observables such as masses, decay constants 
and the $q^2$-behavior of various form factors \cite{Roberts_Dubna,Roberts}. 
The merit of this explicitly  covariant approach is that it takes into 
account the full content of global QCD symmetries from the very beginning.

In the present paper the treatment of the confining interactions 
is based on the widely used instantaneous approximation for
the BS interaction kernel in the c.m. frame of the hadron. This approach,
though fully relativistic, shares most of the simplicity and transparency of
the nonrelativistic model approaches. One may even calculate the corrections
due to the noninstantaneous character of the kernel using the quasipotential
method \cite{Quasi,Brown,Kopal}. Explicit forms of instantaneous $q\bar q$ and
$3q$ kernels are beginning to emerge from lattice simulations \cite{Lattice}
and from QCD-based calculations in the continuum limit \cite{Brambilla,Baker}.
In order to test the kernels it is important to carry out a systematic
quantitative analysis of both the heavy meson and heavy baryon transitions in
the instantaneous picture. Although the effect of the global QCD symmetries
can be consistently embedded in the three-dimensional (3D) approach 
\cite{Kalinovsky}, we do not
deal with a such an extension of the model at the present stage. We restrict
ourselves to the standard relativistic constituent quark model where the
(momentum independent) mass of the constituent quark is the input parameter 
of the theory, rather than that it emerges through the solution of the quark 
Schwinger-Dyson equation.

Most of the recent studies on the BS approach have focused on systematic
investigations of heavy meson weak transitions
\cite{Faustov,Galkin,MesonIW,Huang}.
The calculations in \cite{MesonIW,Huang} are done in the heavy quark limit,
while $1/m_Q$ effects are studied in Refs. \cite{Faustov,Galkin}. In the
baryon sector there has been less activity in the context of the BS approach. 
The reason for this lies in the complexity of the three-body problem both 
technically
and conceptually. Some authors have circumvented the difficulties of the
three-body problem by invoking the quark-diquark picture for the heavy baryons 
which effectively reduces the three-quark system to a two-body bound-state
problem \cite{Guo,Reinhardt}. Summarily one may say that there is ample room
left for systematic studies of heavy baryons as genuine three-body bound
states within the quantum field-theoretical BS approach.

In most of the BS studies of bound state transition amplitudes 
the so-called two-tier scheme is used \cite{Mitra1,Mitra2,Mitra3} when the
instantaneous approximation for the BS kernel is employed. 
In the two-tier scheme one connects 3D
and 4D hadron wave functions according to the following sequence of steps.
First, one reduces the BS equation in the instantaneous approximation to a 
3D equation for the equal-time wave functions and then one solves the 
BS equation. Further, in order
to be able to apply the Mandelstam formalism for the calculation of matrix
elements, one has to "reconstruct" the 4D BS wave function from the equal-time
wave function through the BS equation. The 4D wave function is then substituted
in the resulting expression for the hadronic matrix elements. The two-tier
scheme is well suited for the solution of the two-particle bound state problem.
For the three-particle bound state problem, however, problems arise due to the
{\it disconnectedness} of the three-particle interaction kernel and the choice
of the form of the instantaneous interaction in these disconnected terms.
As a result the final 4D wave function has a rather unusual structure,
containing the square root of the Dirac $\delta$-function. In our approach
such ill-defined structures do not appear.

The aim of the present paper is to calculate the heavy baryon observables in
the covariant instantaneous approximation for the pair-wise kernel of the BS
equation. To this end we develop a framework where the abovementioned problem
related to the ill-definedness of the BS baryon wave function is avoided.
This is achieved by abandoning the two-tier scheme and expressing the matrix
elements directly in terms of the equal-time wave functions following the
ideas of the covariant quasipotential approach \cite{Covariant}. Within our
approach we also calculate the characteristics of heavy meson decays, 
using the same set of parameters as in baryon sector.

As a first approximation to the full complexity of the spin-spin interactions 
we work in the well-known spectator picture 
\cite{DESY,Mitra2,Mitra3,Gudehus,Koerner2}
which provides a well-established setting to include the dynamical effects of
relativity. We shall start, however, from the complete bound-state amplitude 
and outline the approximations which finally lead to the spectator picture. 
In brief, our
approximation consists in expanding the Lorentz-spinor factors in the BS
equation and matrix elements in powers of $|\vec p\,|/m$ and in retaining only
the leading-order term in this expansion (here $|\vec p\,|$ denotes the
magnitude of the relative three-momentum of the quark in the c.m. frame of the
baryon and $m$ stands for its constituent mass). It is obvious that this
approximation differs somewhat from the "static" approximation of
Ref. \cite{Gudehus} which consists in setting all four components of the
quark relative four-momenta equal to zero. We mention, though, that the two
approaches lead to identical results in the analysis of the spin structure
of hadron transition matrix elements. It should be emphasized that the
spectator quark model has been extremely succesful in the description of 
heavy meson and heavy baryon weak transitions \cite{DESY,Gudehus,Koerner2}.
A comprehensive analysis of semileptonic and nonleptonic decay data has been
carried out in this model in terms of a few fit parameters related to the
overlap integrals of the radial part of meson and baryon 4D BS wave functions.
To our knowledge the overlap integrals appearing in the spectator model have
not been calculated yet except the preliminary calculations carried out
within the so-called Lagrangian spectator model \cite{Mainz1,Mainz2}.
One of the aims of our paper consists in establishing a clear and unambiguous
connection of the BS approach in the baryon sector to the spectator quark
model. Such an approach will provide a tool for the microscopic calculation 
of the spectator model parameters as well as for a study of new effects 
beyond the spectator approximation in the heavy hadron weak transitions.

We attempt to remain close to the conventional nonrelativistic
treatment of the bound state problem in terms of 3D equal-time wave functions,
the advantage being that these have a clear physical interpretation.
The merit of such an approach as ours lies in added transparency, and in the
possibility of controlling the magnitude of new relativistic effects.
This can be achieved by restricting the zeroth components of the individual
quark momenta in the baryon equal-time wave functions to their mass shell
values as has been done in most of the studies
\cite{Faustov,Galkin,MesonIW,Huang} in the meson sector. The expressions
obtained for the matrix elements then have a very simple form and can be
readily interpreted in terms of quantum mechanical overlap integrals of 3D
wave functions.

In the present paper we employ the BS framework for heavy-to-heavy transitions
both in the heavy meson and heavy baryon sectors. Using harmonic oscillator
wave functions, we calculate heavy meson and heavy baryon IW functions, their
decay rates and asymmetry parameters. Within the oscillator model for heavy
baryons we also study an alternative approach where the interactions between
the two light quarks in the heavy baryons are switched off. Also, we study 
relativistic effects in the heavy baryon IW function. 

The layout of our paper is as follows: In Sect. 2 we present the BS formalism
for baryons where we discuss in some detail how the instantaneous approximation
can be adapted to the description of heavy baryons. We discuss problems 
related to the disconnectedness of the three-particle BS kernel. 
In Sect. 3 we construct the matrix element of the weak current in terms of 
heavy baryon equal-time wave functions. In Sect. 4 we present the calculation 
of meson and baryon observables in the harmonic oscillator potential model. 
Sect. 5 contains our conclusions.

\section{BS approach to baryons as a bound state system of three quarks}

In this section we shall derive the general BS equation with an instantaneous
kernel for the equal-time baryon wave functions. By taking the limit
$|\vec p\,|/m\rightarrow 0$ in the spinor part we show that the solutions of
the baryonic BS equation reduce to the well-known spectator model wave
functions. We also derive the normalization condition for the equal-time
baryon wave function. The aim of this section is to provide 
equal-time baryon wave functions which can be employed in the
calculation of heavy baryon weak transition matrix elements. A second aim is 
to establish the connection with the quark spectator model wave functions.

\subsection{Bound-state equation}

Let $\Psi_{\alpha\beta\gamma}(p_1p_2p_3)$ denote the 4D BS wave function of the
baryon. We shall express the individual quark momenta $p_i$ through the total
four-momentum
of the baryon ($P$) and the relative Jacobi momenta ($q_1,~q_{23}$) according
to (the $m_i$ stand for the constituent quark masses)
\eq\label{jacobi}
p_1&=&\frac{m_1}{m_1+m_2+m_3}\, P-\frac{1}{3}\, q_1\nonumber\\[2mm]
p_2&=&\frac{m_2}{m_1+m_2+m_3}\, P+\frac{1}{3}\,\frac{m_2}{m_2+m_3}\, q_1
    -\frac{1}{2\sqrt{3}}\, q_{23}\\[2mm]
p_3&=&\frac{m_3}{m_1+m_2+m_3}\, P+\frac{1}{3}\,\frac{m_3}{m_2+m_3}\, q_1
    +\frac{1}{2\sqrt{3}}\, q_{23}\nonumber
\en
The BS equation for the baryon wave function reads
\eq\label{fullBS}
&&(S^{(1)}(p_1))^{-1}\otimes (S^{(2)}(p_2))^{-1}\otimes (S^{(3)}(p_3))^{-1}\,\,
\Psi(p_1p_2p_3)
\nonumber\\[2mm]
&=&\int\frac{d^4k_1}{(2\pi)^4}\frac{d^4k_2}{(2\pi)^4}\frac{d^4k_3}{(2\pi)^4}
\, V(p_1p_2p_3;k_1k_2k_3)\,\,\Psi(k_1k_2k_3)
\en
where $S^{(i)}(p_i)=i(m_i-\not\! p_i)^{-1}$ denotes the propagator of the
i-th quark with momentum $p_i$.
Assuming pair-wise interactions between quarks, the kernel
$V(p_1p_2p_3;k_1k_2k_3)$ in Eq. (\ref{fullBS}) can be written in the
following form
\eq\label{kernel}
V=(2\pi)^4\delta^{(4)}\biggl(\sum_{i=1}^3p_i-\sum_{i=1}^3k_i\biggr)\,
\sum_{cycl~(ijn)}(2\pi)^4\delta^{(4)}(p_i-k_i)\,
(S^{(i)}(p_i))^{-1}\otimes V^{(jn)}(P_{jn};q_{jn},l_{jn})
\en
Here $V^{(jn)}(P_{jn};q_{jn},l_{jn})$ is the two-body potential of the
$(jn)$-th pair and $l_i,~l_{jn}$ denote the relative momenta for the system of
quarks with individual momenta $k_1,~k_2,~k_3$, defined similar to
Eq. (\ref{jacobi}), and $P_{jn}=p_j+p_n=k_j+k_n$.

Next we discuss the choice of the form of the instantaneous kernel in the
three-particle BS equation. For the mesonic two-particle case there exists a
well-established prescription how to obtain the instantaneous kernel. One
constrains the zeroth components of the relative four-momenta of quarks in
the c.m. frame of the meson by the condition $q^0=l^0=0$ which leads to an
instantaneous kernel that depends only on the three-momenta of the quarks.

This procedure cannot be directly generalized to the disconnected
three-particle kernel (Eq. (\ref{kernel})) due to the singular character
of the $\delta$-functions $\delta^{(4)}(p_i-k_i)$ corresponding to
the four-momentum conservation of the spectator quark. In the literature one
finds different prescriptions for the definition of the baryonic 
instantaneous kernel \cite{Mitra1,Mitra2,Mitra3,Dong}. 
The definition of the instantaneous kernel
given in Ref. \cite{Dong}, though natural in view of its nonrelativistic
counterparts, does not possess a natural connection to its relativistic
counterpart given by Eq. (\ref{kernel}). There does not exist a simple
prescription to smoothly extrapolate from kernel (\ref{kernel}) to the
instantaneous limit given in Ref. \cite{Dong}. For this reason we adopt
an alternative definition which was also used in
Refs. \cite{Mitra1,Mitra2,Mitra3}. According to this definition, only
pair-wise interaction kernels $V^{(jn)}$ undergo a (well-defined)
modification in the instantaneous limit. In the c.m. frame of the baryon
the prescription is analogous to that for the two-particle case.
If the instantaneous kernels are assumed to be local, the prescription reads
\eq\label{pair-wise}
V^{(jn)}(P_{jn};q_{jn},l_{jn})\rightarrow
\sum_\Gamma O_\Gamma^{(j)}\otimes O_\Gamma^{(n)}\,\,
V_\Gamma^{(jn)}\biggl(-\frac{\vec q_{jn}-\vec l_{jn}}{2\sqrt{3}}\biggr),
\en
where the matrices $O_\Gamma$ describe the spin structure of the potential
(scalar, vector...). In an arbitrary reference frame the three-vectors are
replaced by the covariant expressions $\vec q_{jn}\rightarrow q_{jn}^T$,
$\vec l_{jn}\rightarrow l_{jn}^T$, where $p_\mu^T=p_\mu-v_\mu(v\cdot p)$ and
$p^\parallel=(v\cdot p)$ etc. (here $v_\mu$ stands for the four-velocity of
the baryon). These substitutions define the transformation rule of the baryon
wave function from the rest frame to an arbitrary frame, providing explicit
Lorentz-covariance of the formalism.

Having chosen the form of the instantaneous interaction, we turn to the
derivation of the bound-state equation where we shall follow the proposals of
Refs. \cite{Mitra1,Mitra2,Mitra3}. We define the equal-time bound-state wave
function according to the conventional prescription \cite{Mitra1}. In the c.m.
frame of the baryon this definition reads
\eq\label{ET}
\tilde\Psi_{\alpha\beta\gamma}(\vec p_1\vec p_2\vec p_3\,)=
\int_{-\infty}^{\infty}\prod_{i=1}^3\,\frac{dp_i^0}{2\pi i}\,\,
\delta\biggl(M_B-\sum_{r=1}^3p_r^0\biggr)\,\Psi_{\alpha\beta\gamma}(p_1p_2p_3)
\en
where $M_B$ denotes the baryon mass. As mentioned before the covariant 
generalization of the above expression is straightforward: $\vec p_i$ is 
replaced by $p_i^T$ and $p_i^0$ by $p_i^\parallel$.

Next, one substitutes the instantaneous kernel defined by Eqs. (\ref{kernel})
and (\ref{pair-wise}) into the BS equation (\ref{fullBS}) and integrates over
relative energy variables. The integral over $q_{jn}^0$ can be easily done 
with the help of the Cauchy's theorem. The remaining integral over $q_i^0$,
however, can not be evaluated by Cauchy's theorem since both the propagators
and the wave function depend on $q_i^0$. In order to proceed one replaces 
$1/3\, q_i^0$ in the propagators by its mass-shell value \cite{Mitra1} 
\eq\label{apprMitra}
1/3\, q_i^0\rightarrow\mu_iM_B-w_i,
\quad\quad\quad
\mu_i=(\sum\limits_{r=1}^3m_r)^{-1}m_i
\en
After integrating over the relative energy variables with the use of the
substitution (\ref{apprMitra}) one can rewrite the BS equation (\ref{fullBS})
to obtain
\eq\label{finishing}
\tilde\Psi(\vec p_1\vec p_2\vec p_3)=
\sum_{i=1}^3\biggl(
\frac{\Lambda_+^{(j)}(\vec p_j)\otimes\Lambda_+^{(n)}(\vec p_n)}
{w_j+w_n+w_i-M_B-i0}
+
\frac{\Lambda_-^{(j)}(\vec p_j)\otimes\Lambda_-^{(n)}(\vec p_n)}
{w_j+w_n-w_i+M_B-i0}
\biggr)\, \hat I^{(i)}\tilde\Psi
\en
where
\eq\label{definitions}
\Lambda_\pm^{(i)}(\vec p_i)=\frac{w_i\pm\hat h_i(\vec p_i)}{2w_i},
\quad\quad
\hat h_i(\vec p_i)=\gamma_0^{(i)}m_i+\gamma_0^{(i)}\vec\gamma^{(i)}\vec p_i,
\quad\quad
w_i=(m_i^2+\vec p_i^{~2})^{1/2}
\en
and
\eq\label{shorthand}
\hat I^{(i)}\tilde\Psi=
\frac{1}{(2\sqrt{3})^3}i\int\frac{d^3\vec l_{jn}}{(2\pi)^3}\,
\sum_\Gamma\,(\gamma_0^{(j)}O_\Gamma^{(j)})\otimes
(\gamma_0^{(n)}O_\Gamma^{(n)})\,\,
V_\Gamma^{(jn)}\biggl(-\frac{\vec q_{jn}-\vec l_{jn}}{2\sqrt{3}}\biggr)
\tilde\Psi(P;\vec q_i,\vec l_{jn})
\en

Eq.(\ref{finishing}) gives a complete set of 3D equations for the equal-time
baryon wave function components, without the need to use a Gordon
on-mass-shell expansion as employed in Refs. \cite{Mitra1,Mitra2,Mitra3}.
The components of the equal-time baryon wave function are defined as
$\tilde\Psi^{\sigma_1\sigma_2\sigma_3}=
\Lambda_{\sigma_1}^{(1)}\Lambda_{\sigma_2}^{(2)}\Lambda_{\sigma_3}^{(3)}
\tilde\Psi$, with $\sigma_1,\sigma_2,\sigma_3=+,-$.
Note that Eq.(\ref{finishing}) differs from the corresponding equation
obtained from the instantaneous kernel used in Ref. \cite{Dong}. Namely,
adopting the kernel given in Ref. \cite{Dong}, it is easy to demonstrate that
only the components $\tilde\Psi^{+++}$ and $\tilde\Psi^{---}$ are nonzero
whereas the "mixed" components present in Eq. (\ref{finishing}) vanish
identically.

In the limit $|\vec p\,|/m\rightarrow 0$ only the $\tilde\Psi^{+++}$-component
of the baryon wave function survives. In this limit Eq. (\ref{finishing}) reads
\eq\label{positive}
\biggl(\sum_{i=1}^3w_i-M_B\biggl)\tilde\Psi^{+++}=
\Lambda_+^{(1)}\Lambda_+^{(2)}\Lambda_+^{(3)}
\sum_{i=1}^3\hat I^{(i)}\,\tilde\Psi^{+++}
\en
Note that the istantaneous kernels of Refs. \cite{Dong} and
\cite{Mitra1,Mitra2,Mitra3} yield the same equation (\ref{positive}) if the
baryon wave function is restricted to the subspace of $(+++)$-components.
Consequently, the difference of our approach to the prescriptions from 
Refs. \cite{Dong} and \cite{Mitra1,Mitra2,Mitra3} reveals itself in the way 
which the $(+++)$-component couples to the negative-frequency components.

\subsection{Wave function}

In the limit $|\vec p\,|/m\rightarrow 0$ in an arbitrary reference frame,
the projectors $\Lambda^{(i)}_\pm$ in Eq. (\ref{definitions}) simplify to
$\Lambda^{(i)}_\pm\rightarrow \frac{1}{2}\,(1+\gamma_0)^{(i)}$
in the c.m. frame of the baryon, or to
$\Lambda^{(i)}_\pm\rightarrow \frac{1}{2}\,(1+\not\! v)^{(i)}$ 
in the general frame. 
In the $|\vec p\,|/m\rightarrow 0$ limit Eq. (\ref{positive}) is solved
by the following ansatz for the wave function
\eq\label{ansatz}
\tilde\Psi_{\alpha\beta\gamma}(p_1^Tp_2^Tp_3^T)=
\theta_{\alpha\beta\gamma}(v)\,\phi(p_1^Tp_2^Tp_3^T)
\en
where $\theta_{\alpha\beta\gamma}(v)$ obeys the following matrix equation
for $(i,j,n)=cycl~(1,2,3)$
\eq\label{cgamma}
\frac{1}{2}\,(1+\not\! v)_{\alpha_1\beta_1}\,\,
\frac{1}{2}\,(1+\not\! v)_{\alpha_2\beta_2}\,\,
\frac{1}{2}\,(1+\not\! v)_{\alpha_3\beta_3}\,\,
\delta_{\beta_i\delta_i}\,\,
(\not\! v O_\Gamma)_{\beta_j\delta_j}
(\not\! v O_\Gamma)_{\beta_n\delta_n}
\theta_{\delta_1\delta_2\delta_3}=
c_\Gamma\theta_{\alpha_1\alpha_2\alpha_3},
\en
The coefficients $c_\Gamma$ are the eigenvalues of the matrix equation 
(\ref{cgamma}). Note that we have written 
$(\gamma_0^{(j)}O_\Gamma)\otimes (\gamma_0^{(n)}O_\Gamma)$ in its covariant 
form $(\not\! v^{(j)}O_\Gamma)\otimes (\not\! v^{(n)}O_\Gamma)$.
Note also that in the limit $|\vec p\,|/m\rightarrow 0$ only the scalar and
zeroth component of vector interactions survive: $O_\Gamma=1,\not\! v$.

In the c.m. frame the radial wave function can be seen to satisfy the
following equation
\eq\label{NRreduced}
\biggl(\sum_{r=1}^3w_r-M_B\biggl)\,\phi(\vec p_1\vec p_2\vec p_3)=
-\frac{1}{(2\sqrt{3})^3}\sum_{i=1}^3\int\frac{d^3\vec l_{jn}}{(2\pi)^3}
u^{(jn)}\biggl(-\frac{\vec q_{jn}-\vec l_{jn}}{2\sqrt{3}}\biggr)
\phi(P;\vec q_i\vec l_{jn})
\en
where $u^{(jn)}=-i\sum\limits_\Gamma c_\Gamma V_\Gamma^{(jn)}$.

The usual ground state baryon spin wave functions can be seen to satisfy
Eq. (\ref{cgamma}) with an eigenvalue $c_\Gamma=1$. Adding the flavor degree
of freedom and putting in the appropriate spin-flavor symmetries one has
(see \cite{Mitra2,Gudehus,Koerner2})
\eq\label{KK}
J^P=1/2^+&\quad&
\theta_{ABC}(v)=
[(\not\! v+1)\gamma^5C]_{\beta\gamma}u_\alpha(v)B_{a[bc]}
+cycl~(a\alpha,b\beta,c\gamma)
\nonumber\\[2mm]
J^P=3/2^+&\quad&
\theta_{ABC}(v)=
[(\not\! v+1)\gamma_\nu C]_{\beta\gamma}u^\nu_\alpha(v)B_{\{ abc\} }
+cycl~(a\alpha,b\beta,c\gamma)
\en
where $B_{a[bc]}$ and $B_{\{ abc\} }$ stand for the flavor wave functions with
a mixed and full symmetry. The index pairs 
$A=(a\alpha),~B=(b,\beta),~C=(c\gamma)$ collect the isospin and Dirac 
indices. Details of the construction of flavor wave functions can be found 
in Refs. \cite{Flavor}.

As a next step one has to specify the pair-wise interaction kernels
$u^{(jn)}$. In the present paper we shall assume that the pair-wise
interactions are of the harmonic oscillator type. One has
\eq\label{ujk}
u^{(jn)}(\vec q-\vec l\,)=\int d^3\vec r\,\,e^{-i(\vec q-\vec l\,)\vec r}\,\,
\biggl(\,\frac{1}{2}\, \mu_{jn}^2\, \omega_0^2\, \vec r^{~2}+u_0^{jn}\biggr),
\quad\quad
\mu_{jn}=\frac{m_jm_n}{m_j+m_n}
\en
Choosing a nonrelativistic form for the quark kinetic energy we obtain 
oscillator wave functions after substituting (\ref{ujk}) in to 
(\ref{NRreduced}).
We present these functions in the c.m. frame of the baryon
($\vec p_1+\vec p_2+\vec p_3=0$). We distinguish between the following cases

\noindent
1. Light baryons containing two quarks of equal mass: $m_2=m_3=m$, $m_1\neq m$
\eq\label{wf1}
&&\phi(\vec p_1\vec p_2\vec p_3\,)
\\[2mm]
&=&C\, \exp\biggl[-\frac{1}{4m\omega_0}
\biggl(\frac{(m_1+m)^{1/2}(m_1+2m)^{1/2}}{m}\, (\vec p_2+\vec p_3\,)^2
+\frac{(m_1+m)^{1/2}}{\sqrt{2}(2m_1+m)^{1/2}}\, (\vec p_2-\vec p_3\,)^2\biggr)
\biggr]
\nonumber
\en

\noindent
2. Heavy-light baryons: $m_1\rightarrow\infty$
\eq\label{wf2}
\phi(\vec p_1\vec p_2\vec p_3\,)=C\, \exp\biggl[-
\frac{1}{2(m_2+m_3)\omega_0}\biggl((\vec p_2+\vec p_3\,)^2+
\frac{(m_3\vec p_2-m_2\vec p_3)^2}{\sqrt{2}m_2m_3}\biggr)\biggr]
\en

\noindent
3. Model of noninteracting light quarks: $u^{(23)}=0$, $m_1\rightarrow\infty$.
The wave function of such a system is remarkably simple since it factorizes
in the variables $\vec p_2$ and $\vec p_3$
\eq\label{wf3}
\phi(\vec p_1\vec p_2\vec p_3\,)=C\,
\exp\biggl[-\frac{\vec p_2^{~2}}{2m_2\omega_0}\biggr]\,
\exp\biggl[-\frac{\vec p_3^{~2}}{2m_3\omega_0}\biggr]
\en
The constant $C$ in Eqs. (\ref{wf1})-(\ref{wf3}) can be determined from the
normalization condition for the wave function (see below). The three 
cases (\ref{wf1}), (\ref{wf2}) and (\ref{wf3}) cover all cases of interest
inasmuch we shall always assume $m_u=m_d$. As has been discussed before,
moving frame wave functions are obtained by the substitution
$\vec p^{~2}\rightarrow -(p^T)^2$.

\subsection{Normalization condition}

In this section we derive the general normalization condition for the
equal-time BS baryon wave function as given in Eq. (\ref{finishing}).
As usual, we start from the 4D BS equation for the six-particle Green's
function $G$ with the instantaneous kernel defined by Eqs. (\ref{kernel}) and
(\ref{pair-wise}). Using Eq. (\ref{apprMitra}), the BS equation can be reduced
to the following 3D equation for the two-time Green's function $\tilde G$
\eq\label{UG}
\tilde G=
-i\tilde g_0\Pi\Gamma_0+i\tilde g_0\sum_{i=1}^3\hat U^{(i)}(M_B)\tilde G
=-i\tilde g_0\Pi\Gamma_0+i\tilde g_0\hat U(M_B)\tilde G
\en
where
$\Gamma_0=\gamma_0^{(1)}\otimes\gamma_0^{(2)}\otimes\gamma_0^{(3)}$ and
\eq\label{def3}
\Pi=\Lambda_+^{(1)}\Lambda_+^{(2)}\Lambda_+^{(3)}
   +\Lambda_-^{(1)}\Lambda_-^{(2)}\Lambda_-^{(3)},
\quad\quad
\tilde g_0=[M_B-\hat h_1(\vec p_1)-\hat h_2(\vec p_2)-\hat h_3(\vec p_3)]^{-1}
\nonumber\\[2mm]
\hat U^{(i)}(M_B)\tilde G=
\tilde g_0^{-1}
\biggl(
\frac{\Lambda_+^{(j)}\Lambda_+^{(n)}}{w_j+w_n+w_i-M_B-i0}
+
\frac{\Lambda_-^{(j)}\Lambda_-^{(n)}}{w_j+w_n-w_i+M_B-i0}
\biggr)\,
\hat I^{(i)}\tilde G
\en
with the operator $\hat I^{(i)}$ given by Eq. (\ref{shorthand}).
From Eq. (\ref{UG}) it immediately follows that
\eq\label{trick}
\tilde G\Gamma_0\Pi\,[\tilde g_0^{-1}-i\hat U(M_B)]\,
\tilde G\Gamma_0\Pi\Gamma_0=
-i\tilde G\Gamma_0\Pi\Gamma_0
\en
Extracting the bound-state pole in the function $\tilde G$, one obtains
\eq\label{contraction}
<\tilde{\bar\Psi}|\Gamma_0\Pi\,[\tilde g_0^{-1}-i\hat U(M_B)]\,|\tilde\Psi>
&=&-(P^2-M_B^2)
\\[2mm]\label{normalization}
<\tilde{\bar\Psi}|\Gamma_0\Pi\,[1-i\frac{\partial}{\partial M_B}\hat U(M_B)]\,|\tilde\Psi>
&=&-2M_B
\en
where $\tilde{\bar\Psi}$ denotes the conjugate wave function. Again the
generalization of the above formulae to an arbitrary reference frame is
straightforward.

Eq. (\ref{normalization}) gives the general normalization condition in the
instantaneous approximation for the baryon equal-time wave function given by 
Eq. (\ref{finishing}). Note that there is an important
difference of the normalization condition (\ref{normalization}) for the
{\it three-particle} wave function as compared to its two-particle counterpart
(see, e.g. \cite{Itzykson}). In the latter case the l.h.s. of the
normalization condition does not depend on the bound-state energy $M_B$ if
the static kernel is energy-independent (a commonly accepted approximation). 
On the other hand the normalization condition for the three-particle wave 
function is nonlinear in $M_B$ irrespective of the form of the potential. 
This energy dependence arises from the energy denominators in Eq. (\ref{def3}).

The normalization condition considerably simplifies in the spectator
approximation. All energy-dependent terms with at least one projector
$\Lambda_-^{(i)}$ drop out in this limit. With the help of Eq. (\ref{ansatz}),
one concludes from Eq. (\ref{normalization}) that
\eq\label{norm_S}
N_C!\bar\theta\theta\,\,\frac{1}{(6\sqrt{3})^3}\int
\frac{d^3\vec q_1}{(2\pi)^3}\frac{d^3\vec q_{23}}{(2\pi)^3}\,\,
\phi^2(\vec q_1\vec q_{23})=2M_B
\en
where the factor $N_C!=3!$ arises from the sum over (implicit) color indices
and
\eq
\bar\theta\theta=\bar\theta_{\alpha\beta\gamma}\theta_{\alpha\beta\gamma},
\quad\quad
\bar\theta_{\alpha\beta\gamma}=\theta^\star_{\alpha'\beta'\gamma'}
(\gamma_0)_{\alpha'\alpha}(\gamma_0)_{\beta'\beta}(\gamma_0)_{\gamma'\gamma}
\en
Using Eq. (\ref{norm_S}) and the explicit expressions for the oscillator
wave functions, Eqs. (\ref{wf1})-(\ref{wf3}) it is then a simple task
to calculate the normalization factor $C$. One has 

\noindent
1. Light baryons containing two quarks of equal mass: $m_2=m_3=m$, $m_1\neq m$
\eq
C=2^{13/8}\pi^{3/2}\,
\biggl(\frac{M_B}{N_C!\bar\theta\theta m^3\omega_0^3}\biggr)^{1/2}
\biggl(\frac{m_1+m}{m_1}\biggr)^{3/4}
\biggl(\frac{2m_1+4m}{2m_1+m}\biggr)^{3/8}
\en

\noindent
2. Heavy-light baryons: $m_1\rightarrow\infty$
\eq
C=2^{13/8}\pi^{3/2}\,
\biggl(\frac{M_B}{N_C!\bar\theta\theta (m_2m_3)^{3/2}\omega_0^3}\biggr)^{1/2}
\en

\noindent
3. Two noninteracting light quarks
\eq
C=4\pi^{3/2}\,
\biggl(\frac{M_B}{N_C!\bar\theta\theta (m_2m_3)^{3/2}\omega_0^3}\biggr)^{1/2}
\en

\section{Matrix elements of heavy baryon transitions}

Below we give the expression of the matrix element of the weak current
$\bar Q^\prime(0)W^\mu Q(0)$ with $W^\mu=\gamma^\mu(1+\gamma^5)$
between heavy baryon states. In the derivation we follow the ideas of the
covariant quasipotential approach \cite{Covariant}. However, there is a
difference between our approach and the commonly used quasipotential
approaches related to the treatment of the negative-frequency components of
the baryon wave function. As is well known for the case of spin-$\frac{1}{2}$
constituents, neither the free nor the full equal-time Green's function can be 
inverted. In the quasipotential method the free Green's function is modified
such that it can be inverted. The requisite modifications are by no means
unique and differ by how the negative-frequency components of the wave
function are treated. In general this may lead to different results for 
heavy baryon transition matrix elements. In our approach there is no need 
for such a modification and one can retain the full content of the equal-time 
BS wave function frequency components in the matrix elements.

After these introductory remarks we turn to the derivation of the matrix
elements in the BS approach within the spectator model approximation. We 
denote the heavy quark ($c$ or $b$) in the baryon by the label $1(1')$, 
while labels $2$ and $3$
correspond to the light quark constituents. For the time being we keep the
mass of the heavy quark finite. Let $R^\mu$ denote the Green's function for
the $Qqq\rightarrow Q^\prime qq$ transition induced by the weak current.
In the lowest-order approximation $R^\mu$ reads
\eq\label{pspaceRIA}
R^\mu=S^{(1')}(p_1)W^\mu S^{(1)}(k_1)\otimes
(2\pi)^4\delta^{(4)}(p_2-k_2) S^{(2)}(p_2)\otimes
(2\pi)^4\delta^{(4)}(p_3-k_3) S^{(3)}(p_3)
\nonumber
\en
The two-time operator $\tilde R^\mu$ is defined by
\eq
&&\tilde R^\mu(v',v)=
\int_{-\infty}^{\infty}\prod_{i=1}^3\frac{da_i}{2\pi i}\frac{db_i}{2\pi i}\,\,
2\pi i\delta\biggl( M'_B-\sum_{r=1}^3a_r\biggr)\,
2\pi i\delta\biggl( M_B-\sum_{s=1}^3b_s\biggr)\,\,
R^\mu(p_1p_2p_3;k_1k_2k_3)
\nonumber\\[2mm]
&&a_i=(v^\prime\cdot p_i),
\quad\quad
b_i=(v\cdot k_i)
\en
where $v(v')$ and $M_B(M'_B)$ denote the velocity and the mass of the
initial (final) baryon.

In the conventional relativistic impulse approximation one takes
the lowest-order result (\ref{pspaceRIA}) and neglects the interaction
terms $-i\hat U$ in Eq. (\ref{contraction}).
Extracting the double pole in $R^\mu$ with the use of
Eqs. (\ref{contraction}) and (\ref{normalization}) one obtains the following
expression for the matrix element of the weak current 

\eq\label{sought-for}
<P^\prime|\bar Q^\prime(0)W^\mu Q(0)|P>=
-<\tilde{\bar\Psi}_v^\prime|
\Gamma_0(v)\Pi(v^\prime)
\tilde g_0^{-1}(v^\prime)
\tilde R^\mu(v',v)
\Gamma_0(v')\Pi(v)
\tilde g_0^{-1}(v)
|\tilde\Psi_{v}>
\en
where all internal integrations are three-dimensional. In the
spectator approximation the general structure of the transition matrix
element (\ref{sought-for}) can be seen to further simplifies. The details 
of the derivation can be found in the Appendix. The final result reads
\eq\label{element}
<P^\prime|\bar Q^\prime(0)W^\mu Q(0)|P>=2(M_BM^\prime_B)^{1/2}
(\bar\theta\theta)^{-1}\,\,
\bar\theta(v^\prime)W^\mu\theta(v)\,\,f(v\cdot v^\prime)
\en
with
\eq\label{fomega}
&&f(v\cdot v^\prime)=\frac{N_C!\bar\theta\theta}{2(M_BM^\prime_B)^{1/2}}\,\,
\frac{1}{(6\sqrt{3})^8}\,\int
\frac{d^4q_1}{(2\pi)^4}\frac{d^4q_{23}}{(2\pi)^4}
\frac{d^4l_1}{(2\pi)^4}\frac{d^4l_{23}}{(2\pi)^4}\,\,
(2\pi)^4\delta^{(4)}(\Delta_2)(2\pi)^4\delta^{(4)}(\Delta_3)
\nonumber\\[2mm]
&\times&
F(\vec q_1\vec q_{23})\bar S_H^{(1')}(p_1^{\prime 0})
\bar S_H^{(1)}(k_1^{\prime 0}) F(\vec l_1\vec l_{23})\,\,
\bar S_L^{(2)}(p_2')\bar S_L^{(3)}(p_3')
\en
where $F(\vec p_1\vec p_{23})=(\bar\Lambda-w_2-w_3)\phi(\vec p_1\vec p_{23})$, 
$M_B=m_1+\bar\Lambda+O({m_1}^{~-1})$ and  
$M'_B=m'_1+\bar\Lambda+O({m'_1}^{~-1})$. The four-vectors $p_i'$, $k_i'$
are defined by Eqs. (\ref{momentum-prime}) in Appendix. The propagators
in the spectator approximation are given by the following expressions
\eq
\bar S_H(p^0)=\frac{1}{-\bar\Lambda+m_2+m_3-p^0-i0}
&\hspace*{1cm}& \mbox{for~heavy~quarks}\nonumber\\[2mm]
\bar S_L^{(i)}(p)=\frac{1}{w'_i-m_i-p^0-i0}
&\hspace*{1cm}& \mbox{for~light~quarks}
\en
For the light quarks ($i=2,3$) one has
\eq
\Delta_i^0&=&m_i({v'}^{0}-v^0)+{{p_i'}^0}{v'}^{0}+{\vec p_i^{~\prime}}
\vec v^{~\prime} -{{k_i'}^0}{v}^{0}-{\vec k_i~'}\vec v
\\[2mm]
\vec\Delta_i&=&m_i(\vec v^{~\prime}-\vec v\,)
+{\vec p_i~'}+{{p_i'}^0}\vec v^{~\prime}
+(v^{\prime 0}+1)^{-1}({\vec p_i~'}\,\vec v^{~\prime}\,)\vec v^{~\prime}
-{\vec k_i~'}+{{k_i'}^0}\vec v
+({v}^{0}+1)^{-1}({\vec k_i~'}\,\vec v\,)\vec v
\nonumber
\en

The Lorentz structure of the current matrix element 
Eq. (\ref{element}) is determined by the spectator model factor
$\bar\theta(v')W^\mu\theta(v)$. It is well known that in the heavy quark limit
baryonic ground-state to ground-state transitions are determined by three
independent form factor functions $\zeta(\omega)$, $\xi_1(\omega)$ and
$\xi_2(\omega)$ which depend ones on the momentum transfer variable
$\omega=v\cdot v'$ \cite{Isgur-Georgi}. In the spectator model these three
functions become related and are given in terms of a single universal form
factor function $f(\omega)$ \cite{DESY}. One has
\eq\label{BarIW}
\zeta(\omega)=\xi_1(\omega)=\xi_2(\omega)(\omega+1)=
f(\omega)\frac{\omega+1}{2},
\quad\quad
f(1)=1
\en
\noindent This result coincides with the prediction of large-N$_c$ QCD
\cite{Chow}.

In order to determine the reduced form factor function $f(\omega)$, it is
sufficient to consider only one particular transition. For example, take the
$\Lambda_b\to\Lambda_c$ transition. Using the known spectator model wave
functions Eq. (\ref{KK}), one obtains
\eq\label{spec-lambda}
\bar\theta_{\Lambda_c}(v^\prime)W^\mu\theta_{\Lambda_b}(v)=
\frac{1}{2}\bar\theta_{\Lambda_c}\theta_{\Lambda_b}
(1+\omega)\bar u(v^\prime)W^\mu u(v)
\en
Then from Eqs. (\ref{element}) and (\ref{spec-lambda}) one immediately
concludes that
\eq\label{add-new}
<\Lambda_c|\bar Q^\prime(0)W^\mu Q(0)|\Lambda_b>=
2(M_{\Lambda_c}M_{\Lambda_b})^{1/2}
\bar u(v^\prime)\gamma^\mu(1+\gamma^5)u(v)\,\,\frac{\omega+1}{2}\,f(\omega)
\en
From Eq. (\ref{add-new}) it is seen that the universal function $f(\omega)$
in Eq. (\ref{BarIW}) coincides with the one given by Eq. (\ref{fomega}).
Using the normalization condition for the baryon wave function (\ref{norm_S}),
one can check that normalization condition $f(1)=1$ is satisfied.

In order to proceed further in the calculation of the heavy baryon weak
semileptonic transition matrix element given by Eqs. (\ref{element}) and
(\ref{fomega}), we choose a particular reference frame where
$v^\mu=(1,0,0,0)$ and $v^{\prime\mu}=(\omega,0,0,\shw)$. After integrating
over the variables $l_1$ and $l_{23}$ the arguments of the initial wave
function and $w_1,~w_2$ become dependent on the relative energy variables
$q_1^0$ and $q_{23}^0$. Cauchy's theorem can therefore not be used in the
evaluation of the integrals over $q_1^0$ and $q_{23}^0$. As mentioned before 
this is similar to that happens in the mesonic case. 
Also this dependence gives rise to a spurious imaginary part
in the function $f(\omega)$ at $\omega\neq 1$. A simple way to remedy this
difficulty is to fix the relative energies on mass shell (the same, at the
poles of the denominator in the Eq. (\ref{fomega})) in the wave functions
and in the quantities $w_1,~w_2$ such that one has
\eq
\frac{1}{3}\, q_1^0=w'_2+w'_3-m_2-m_3,
\quad\quad
\frac{1}{2\sqrt{3}}\, q_{23}^0=\frac{m_2w'_3-m_3w'_2}{m_2+m_3}
\en
The Cauchy integral over the energy denominators can then easily be performed.
The factor $(\bar\Lambda -w_2-w_3)(\bar\Lambda -w'_2-w'_3)$ in the numerator
is cancelled upon integration and we are left with the simple result
\eq\label{overlap}
f(\omega)=\frac{N_C!\bar\theta\theta}{2(M_BM'_B)^{1/2}}\frac{1}{(6\sqrt{3})^3}
\int\frac{d^3\vec q_1}{(2\pi)^3}\frac{d^3\vec q_{23}}{(2\pi)^3}\,\,
\phi(\vec q_1\vec q_{23})\phi(\vec l_1\vec l_{23})
\en
where
\eq\label{boost-relative}
&&l_1^3=3\shw(w'_2+w'_3)+\omega q_1^3,
\quad
l_{23}^3=\omega q_{23}^3+\shw 2\sqrt{3}\,\frac{m_2w'_3-m_3w'_2}{m_2+m_3},
\quad
\! \! \vec l_1^{~\bot}=\vec q_1^{~\bot}
\nonumber\\
&&
\en
The physical meaning of the result is transparent. Eq. (\ref{overlap})
corresponds to the quantum-mechanical overlap of two baryon wave functions.
The initial wave function is evaluated in the rest frame of the initial baryon
and the final wave function in the frame moving with the velocity $v'$ along
the third axis. The arguments of the final wave function are Lorentz boosted
where the energies of the light quarks are fixed by their mass shell values.
Obviously, the same result can be obtained from the general expression
(\ref{sought-for}) e.g. in the rest frame of the final baryon. In this case
the initial wave function in Eq. (\ref{overlap}) is substituted by the final
wave function and vice versa. Since in the heavy quark limit the wave
functions do not depend on heavy flavor, one ends up with the same heavy
baryon IW function in both frames.

The calculation of the heavy meson IW function proceeds along similar lines
but will not be presented in this paper. We only give the final result
obtained with the same assumptions as for the case of baryons. For the reduced
form factor function one obtains
\eq\label{meson_IWF}
\xi(\omega)=\int\frac{d^3\vec q}{(2\pi)^3}\phi_M(\vec q\,)\phi_M(\vec l\,),
\quad\quad
\vec l^{~\bot}=\vec q^{~\bot},
\quad\quad
\l^3=\omega q^3+\shw (m^2+\vec q^{~2})^{1/2}
\en
where the ET meson wave function $\phi_M(\vec q\,)$ is normalized to unity,
i.e. one has the normalization $\xi(1)=1$.

It is interesting to note that in the approximation of noninteracting
light quarks within the heavy baryon the meson and baryon IW functions
become related if one assumes that the interaction potentials between
the heavy quark and the light quark/antiquark are the same
\cite{Koerner-triangle}. Let $\zeta(\omega)$ be the IW function describing
the transition $\Lambda_b\to\Lambda_c$ and $\xi(\omega)$ the mesonic IW 
function. Using Eq. (\ref{overlap}) one then obtains
\eq\label{factorization}
\zeta(\omega)=\frac{1}{2}(\omega+1)f(\omega)=
\frac{1}{2}(\omega+1)\xi^2(\omega)
\en
The two light quarks in the heavy baryon  
move independently in the mean field produced by the heavy quark where the 
heavy quark is fixed in the center of mass of the heavy baryon. Such a 
physical picture is quite attractive since one can relate the heavy baryon form
factors to the heavy meson form factors (to be more precise, to the would-be 
heavy meson form factors, in which the interquark interaction potential 
coincides, by definition, with the potential acting between the heavy and 
light quarks in the heavy baryons). Anyway, the model of noninteracting light 
quarks enables one to effectively reduce the calculation of heavy baryon 
observables to the two-body case and thus enormously simplifies the treatment 
of the problem under study.

The assumption of noninteracting light quarks has two aspects which one may 
refer to as "kinematical" and "dynamical" aspects. Let us elaborate on these 
two aspects. 
The kinematical aspect deals with the spins of the quarks and manifests
itself in the relativistic factor $\frac{1}{2}(\omega+1)$ in e.g.
Eq. (\ref{factorization}). We would like to emphasize that the kinematical
aspect of the noninteracting light quark model is already implicit in the
spectator model wave functions which are derived from the equal-velocity
assumption (all quarks being on mass shell and propagating freely). Not
surprisingly, the overlap integral of the baryon wave functions contains
the factor $\frac{1}{2}(\omega+1)$ explicitly (see e.g.
Eq. (\ref{spec-lambda})). The physical origin of this factor can be seen
by considering the transition amplitude in the crossed channel, corresponding
to the production of the heavy baryon-heavy antibaryon pair by the virtual
photon born in the $e^+e^-$ annihilation process.

Let us first consider the physical picture where both light quarks are
produced independently from the vacuum through the exchange of many soft
gluons with the total quantum numbers $J^P=0^+$ (Fig. 1a). The intrinsic
parity of the $\biggl(\frac{1}{2}^+\overline{\frac{1}{2}^+}\biggr)$ pair
is negative and, consequently, $(LSJ)=(110)$ for this transition.
Thus one has a threshold factor of $|\vec p\,|$ for each of the two $P$-wave 
quark-antiquark pairs, i.e. in total one has a $|\vec p\,|^2$ threshold 
factor where $|\vec p\,|$ is the magnitude of the c.m. relative
three-momentum of the quark pair.

As opposed to this, let us consider the situation when two light quarks
inside the heavy baryon are tightly bound in a diquark with the quantum
numbers $J^P=0^+$ (Fig. 1b). Then the amplitude for the transition
$0^+\to 0^++\overline{0^+}$ is an $S$-wave transition without 
any threshold factor.

In the equal-velocity approximation the magnitude of the
c.m. relative three-momentum can be expressed in terms of the velocity
transfer variable $\omega=(v\cdot v')=M_B^{-2}(p_1\cdot p_2)$ where
$p_1$ and $p_2$ are the momenta of baryon and antibaryon
produced in the $e^+e^-$ annihilation. It is a simple task to derive
$|\vec p\,|/m=|\vec v~'|=\biggl(\frac{1}{2}(\omega-1)\biggr)^{1/2}$.
In the direct channel $\omega$ is replaced by $-\omega$ and thus the threshold
factor $|\vec p\,|^2$ turns into $\frac{1}{2}(\omega+1)$ present in
Eq. (\ref{spec-lambda}).

The dynamical aspect of the noninteracting light quark model consists in the
assumption of the factorization of baryon radial wave function with regard to 
variables $\vec p_1$ and $\vec p_2$. This can be achieved by setting the
interaction potential between the light quarks to zero. From a 
rigorous point of view, one should then also replace the interaction 
between the heavy and light quarks in the heavy baryon by some
effective "mean field" interaction. Below we shall present the results of
numerical calculations which demonstrate the effect of the noninteracting
light quark approximation in heavy baryon observables.

\section{Results}

In this section we present our numerical results both for heavy meson and
heavy baryon sectors. We use oscillator wave functions as given by 
Eqs. (\ref{wf1})-(\ref{wf3}) for baryons and corresponding oscillator wave
functions for mesons. Oscillator wave functions are known to provide a good 
basis of trial wave functions in the variational solution of the bound-state
equation \cite{Kopal,Richard}.

We would like to emphasize that in the present paper we have not attempted
to obtain a precise description of meson and baryon data by the fine
tuning of a large number of model parameters. Rather, we want to demonstrate 
that, in the framework considered in the present paper, one achieves 
a reasonably good description of experimental numbers both in the meson and 
baryon sector with only a few parameters. We have also checked that the 
dependence of our numerical results on these parameters is rather moderate. 

In order to reduce the number of free parameters as much as possible we do not
distinguish between the masses of the light quarks and set $m_u=m_d=m_s=m$. 
It is known that the effect of the $m_s-m_u$ mass difference in the baryon 
wave functions is rather small and we neglect it in the present treatment.

First we turn to the calculation of the heavy meson leptonic decay constants
defined by
\eq\label{leptonic-constant}
iP_\mu f_P=-iN_C\int\frac{d^3\vec p}{(2\pi)^3}\,\,
{\rm Tr}[\tilde\chi(P;\vec p\,)\gamma_\mu\gamma^5]
\en
where $\tilde\chi(P;\vec p\,)$ denotes the equal-time meson wave function in
the c.m. frame. In an arbitrary reference frame the heavy meson wave function
is given by (see e.g. \cite{Koerner2})
\eq
\tilde\chi(P;p^T)=c_M\gamma^5(1-\not\! v)\,M_a^b\,\phi_M[-(p^T)^2]
\en
Here $M_a^b$ denotes the meson flavor matrix and $c_M$ is the normalization
constant. Using the BS normalization condition and assuming the radial part
of the meson wave function to be of the oscillator type
$\phi_M\sim\exp[(p^T)^2/\Lambda^2]$, it is a straightforward task to obtain
\eq\label{fp}
%f_P=2^{1/2}\pi^{-3/4}N_C^{1/2}M_P^{-1/2}\Lambda^{3/2}
f_P=\biggl(\frac{2N_C\Lambda^3}{\pi^{3/2}M_P}\biggr)^{1/2}
\en
where $M_P$ denotes the meson mass. As can be seen from Eq. (\ref{fp}) 
the calculated leptonic decay constant exibits the well-known $M_P^{-1/2}$
scaling behavior.

Next we turn to the calculation of the heavy meson Isgur-Wise (IW) function 
according to Eq. (\ref{meson_IWF}). The IW function depends only on the ratio 
$m/\Lambda$ where $m$ is the light quark mass. 
In order to fix this ratio we calculate the slope parameter of the heavy 
meson's IW wave function using again oscillator wave functions for the heavy 
meson. One obtains 
\eq\label{meson_r2}
\rho^2=\frac{3}{4}+\frac{m^2}{\Lambda^2}
\en
We further use the calculated values of leptonic decay constants in 
Eq. (\ref{fp}) to provide absolute values for $m$ and $\Lambda$. 
Most of the present theoretical investigations of the slope parameter 
converge around the value $\rho^2\approx 1$. With $\rho^2=1$ as input 
one obtains $\Lambda=2m$ from Eq. (\ref{meson_r2}). Further, taking 
the value $m=250$~MeV
for the constituent quark mass and, respectively, $\Lambda=500$~MeV for the
wave function range parameter, we obtain a reasonable fit to the experimental 
leptonic decay constants as shown in Table I. For comparison, 
we have also
listed the results of the recent lattice calculations of the same quantities 
in Table I. 
In Table II we give some recent results on the heavy meson IW function slope
parameter. Note that the functional dependence of the heavy meson IW function 
in our approach is well approximated by the formula
\eq\label{parameterization}
\xi(\omega)=\biggl(\frac{2}{\omega+1}\biggr)^{2\rho^2}
\en
in our approach where, as was mentioned above, we take $\rho^2=1$ as input.

With the above two parameter values we present our results
of the calculation of the decay observables in $B$-meson semileptonic
transitions in TABLE III. We give the branching ratios for the weak 
semileptonic decays
$B\to D,~B\to D^\star$ and values for the polarization-type observables
$\alpha_{pol},~\alpha',~A_{FB}~\mbox{and}~A_{FB}^T$ \cite{Neubert,Bialas}
in these decays. As can be seen from table III, the agreement of the
calculated quantities with existing experimental data is satisfactory.

Next we present our results in the baryon
sector. We use the same values for the two model parameters
as in the meson sector. To begin with we discuss the model of
noninteracting light quarks with the additional assumption that the interaction
between the heavy and light quarks in the heavy baryon is the same as the
interaction between the heavy quark and light antiquark in the heavy meson
\cite{Koerner-triangle}. With this assumption the range parameter in the 
heavy baryon
equal-time wave function defined by $\Lambda_B=2m\omega_0$ turns out to be the
same as the parameter $\Lambda$ in the meson wave function. Consequently, the
heavy baryon IW function can be also parameterized by the ansatz
(\ref{parameterization}) but with
\eq\label{baryon_r2}
\rho^2_B=1+\sum_{light}\biggl(\frac{m_{light}}{\Lambda}\biggr)^2
=2\rho^2-\frac{1}{2}
\en
such that $\rho^2_B=1.5$ in the model of noninteracting light quarks.

When we present our results in the following on the functional dependence of 
the heavy baryon IW function $\zeta(\omega)$ we shall always compare the two 
cases where the interaction between the light quarks is switched either on
or off. The results of our calculation are given in Fig. 2. As can be seen  
from Fig. 2 the IW function is substantially flatter (with the same set of 
model parameters $m/\Lambda=0.5$) when the interactions in the light 
diquark are taken into account. At maximum recoil the difference between the 
values of the function $\zeta(\omega)$ calculated in the full (interacting) 
model and in the noninteracting light quark model amounts up to $30~\%$. 
The function $\zeta(\omega)$ in the full model is well approximated by 
the functional dependence 
\eq\label{zeta-approx}
\zeta(\omega)=\biggl(\frac{2}{\omega+1}\biggr)^{a+b/\omega}
\en
with $a=1.23$ and $b=0.4$. This corresponds to a slope parameter of 
$\rho_B^2=0.81$ which is much lower than the slope parameter $\rho_B^2=1.5$ 
in the noninteracting light quark model.

In our simplified approach the heavy baryon IW function depends only on the 
ratio $m/\Lambda_B$. Fixing $m$ at $m=250$ MeV we have evaluated the IW 
function 
for two different values of $\Lambda_B$. In Fig. 2 we present our results 
for the value $\Lambda_B=\Lambda/\sqrt{2}=355$~MeV, 
corresponding to the popular one half rule for baryons 
(see e.g. \cite{Richard}). According to this
rule which is strongly supported by phenomenology, the interactions between
quarks in the baryon are down by the factor $\frac{1}{2}$ as compared to the
interactions between the quark and antiquark in the meson. From Fig. 2 one 
observes that even for such a substantial change of the value of the
parameter $\Lambda_B$ the IW function is only sligthly modified. 
In particular the difference between the noninteracting and full calculation 
persists. The
maximum velocity transfer the IW functions calculated in the full model with
these two values of $\Lambda_B$ differ only by about $5~\%$. The functional
dependence of the IW function with $\Lambda_B=355$~MeV can again be well 
approximated by the representation (\ref{zeta-approx}) but now with $a=1.54$, 
$b=0.4$; $\rho_B^2=0.97$.

To summarize our results on the functional dependence of the Isgur-Wise 
function one finds that the model of noninteracting light quarks can be 
clearly distinguished 
from the full model, at least within the spectator picture. There is no 
theoretical reason to prefer one to the other. The issue which of the 
two ans\"{a}tze has to preferred has to be settled by experiments. An
attractive feature of our approach, apart from its simplicity, consists in the
weak dependence of the predictions on the precise values of the model 
parameters which helps in distinguishing between various models.

In Table IV we present the calculated total widths of bottom baryon weak
semileptonic transitions calculated in the full model with $\Lambda_B=500$~MeV,
column (1), and $\Lambda=355$~MeV, column (2). For comparison we also list 
results from other model approaches. We see that the overall agreement of our 
results with results from other models discussed in the literature is 
reasonable. In the absence of experimental data, however,
one can not fix the precise value of the wave function range parameter
$\Lambda_B$ owing to the weak dependence of the results on this parameter.

In Table V we give values for the asymmetry parameters in the $\Lambda_b$ 
baryon semileptonic transitions \cite{Ivanov,Koerner-PLB}. Rows (1) and (2) 
are for $\Lambda_B=500$ MeV and $\Lambda_B=355$ MeV as in Table IV. For 
comparison, we again list the
results obtained within different approaches. As can be seen from Table V 
the asymmetry parameters are rather insensitive to the particular model in
which they are calculated.

Last but not least, we shall discuss the nonrelativistic limit of our approach
where the crucial role of the factor $\frac{1}{2}(\omega+1)$ will become 
transparent. First, note that for
$v^{\mu}=(1,\vec 0\,)~\omega=v^{\prime 0}=
\sqrt{1+\vec v^{~\prime 2}}\approx 1$
up to the terms of order of $\vec v^{~\prime 2}$. Thus this factor drops in the
nonrelativistic limit. Further, from Eqs. (\ref{boost-relative}) 
one obtains
\eq\label{momenta-NR}
\vec l_1=\vec q_1+6m\vec v^{~\prime},
\quad\quad\quad
\vec l_{23}=\vec q_{23}
\en
in this limit 
which coincides with the formulae presented in Refs. \cite{Singleton,Cheng}.
From Eqs. (\ref{momenta-NR}) one immediately obtains 
the nonrelativistic (NR) result for the Isgur-Wise function 
\eq\label{IW-NR}
\zeta_{NR}(\omega)=f_{NR}(\omega)=
\exp\biggl[-\frac{m^2}{\Lambda^2}(\omega^2-1)\biggr]
\en
with $\rho_{B,NR}^2=0.5$.
In Fig. 3 we have plotted the functions $f(\omega)$ and
$\zeta(\omega)$ in the full model in the relativistic case as well as the
function $\zeta_{NR}(\omega)$ given by Eq. (\ref{IW-NR}). It is interesting to
note that although there exists a large relativistic dynamical effect since 
$\zeta_{NR}(\omega)$ and $f(\omega)$ are significantly different, the absence
of the relativistic factor $\frac{1}{2}(\omega+1)$ in the nonrelativistic case
compensates for this. In the heavy meson case where such 
a factor is absent the slope parameter $\rho^2$ of the IW function is grossly
underestimated in the nonrelativistic treatment \cite{Close,LeYaouanc}.
In our opinion our results unambigouosly indicate the
importance of including relativistic effects when studying heavy
baryon transitions.

\section{Summary and outlook}

We have presented a simple calculation of the heavy meson
and heavy baryon semileptonic decay observables with the
use of the field-theoretical BS approach where both heavy mesons and heavy 
baryons are treated on the same footing. While widely in use in the 
calculation of 
mesonic two-body bound-state observables, the BS approach has been known to 
encounter conceptual difficulties when applied to the baryonic 
three-body  case. In the present paper we have explicitly demonstrated
that the treatment of the three-body bound state systems 
can proceed along similar lines as in the two-body case 
when the constituents interact via instantaneous kernels. 

Up to this point our investigation of the three-body problem has been 
restricted to the so-called "spectator picture" which provides a powerful 
tool for the study of heavy baryon weak interactions. In the spectator 
approximation the spin structure of the baryon wave functions and decay 
matrix elements is
remarkably simple, and model-independent relations emerge between various
decay amplitudes in this limit. This has been demonstrated by a systematic 
and comprehensive analysis of the heavy baryon weak
nonleptonic decays, carried out in papers \cite{Gudehus,Koerner2}. Further
improvements on these ideas were given in Refs. \cite{Mainz1,Mainz2}, where the
so-called Lagrangian spectator model has been proposed. The Lagrangian 
spectator model allows for the microscopic evaluation of the various 
"overlap integrals" (the reduced matrix elements, from the group-theoretical 
point of view), and thus allows one to compute a large number of experimental 
observables in heavy baryon decays: rates, asymmetry parameters, etc.

The spectator approximation is based upon a very simple and transparent
physical picture: the internal motion of quarks (both heavy and light) inside
the hadron is very slow; all quarks are assumed to be on mass shell and are 
assumed to have the same velocity, which coincide with the velocity of the 
hadron as a whole. All approximations which we have used in the treatment of
the heavy baryon weak transitions are in accordance with the above physical
picture, and can be deduced from it. Indeed, in the present paper we have
demonstrated that the general BS approach to transition matrix elements can 
be reduced, step-by-step, to the spectator model treatment of the weak 
transition matrix elements with the use of the above approximations. 
We would like to emphasize that the BS approach enables one to evaluate 
the full  matrix elements by expressing them in terms of the overlaps between 
equal-time BS wave functions. Note also, that
even in the spectator limit our approach is {\it not} reduced to a 
nonrelativistic approach: only that part of motion which is related to the 
quark relative momenta is treated nonrelativistically, whereas the c.m. 
motion of hadrons is taken into account in a completely relativistic fashion. 
Thus,
even at this stage our model is not reduced (and differs significantly) from
the nonrelativistic baryon models which are used for the calculation of
transition amplitudes \cite{Singleton,Cheng}. Moreover, the necessity of the
inclusion of the relativistic effects is readily seen even from the results
of our calculations in the spectator picture.

We would like to mention that the physical assumption of "slow" interquark
motion for the light quarks (which lies at the heart of the whole spectator 
picture) cannot be rigorously justified from a theoretical point of view. 
According to common belief one has the value $|\vec p\,|/m\sim 1$ for
light quarks inside the hadron. Since, the quark spin-spin interaction
effects are proportional to powers of $|\vec p\,|/m$ they are
{\it a priori} expected to give a sizeable contribution to the calculated
baryon observables in contrast to the assumptions of the spectator model. 
Nevertheless, the present treatment of 
heavy hadron transitions is based on a relativistically consistent 
formulation of 
the spectator picture. It can be used as a stepping stone for the inclusion 
of $O(|\vec p\,|/m)$ spin effects at a later stage. We plan
to address this problem in future publications.

In addition, we plan to apply the present BS approach and its possible
modification with the inclusion of the spin effects to the more involved
and interesting problems of heavy hadron physics, such as nonleptonic, 
one-pion and radiative decays of heavy baryons.

\vspace*{.3cm}

\noindent
{\it Acknowledgments}

\vspace*{.3cm}
\noindent
M.A.I, V.E.L and A.G.R thank Mainz University for the hospitality
where a part of this work was completed.
A.G.R. is thankful to Prof. T. Kopaleishvili for the introduction
to the baryon BS equation.
This work was supported in part by the Heisenberg-Landau Program,
by the Russian Fund of Basic Research (RFBR) under contract
96-02-17435-a and by the BMBF (Germany) under contract 06MZ566.
V.E.L. thanks the Russian Federal Program ``Integration of Education and 
Fundamental Science'' for partial support. 

\vspace*{2.cm}

\appendix

\section{Matrix elements in the spectator approximation}
\def\theequation{A.\arabic{equation}}
\setcounter{equation}{0}

In this Appendix we present details of how to evaluate the heavy baryon weak 
transition matrix elements in the spectator approximation. In particular, 
we shall
demonstrate that the spin structure of Eq. (\ref{sought-for}) considerably
simplifies in the spectator picture. First, note that the factors
$\Gamma_0$ and $\Pi$ in Eq. (\ref{sought-for})can be dropped since they are 
reduced to an identity operator when acting on the spectator model wave 
functions. Moreover, $\tilde g_0^{-1}$ is reduced to
\eq
\tilde g_0^{-1}(v')\rightarrow
M_B-\sum_{i=1}^3(m_i^2-p_i^2+(p_i\cdot v')^2)^{1/2}
\en
An analogous relation holds for $\tilde g_0^{-1}(v)$.

In order to further simplify Eq. (\ref{sought-for}) we perform a change
of integration variables corresponding to Lorentz boosts which boost 
the initial and final baryon wave functions to the rest frame. One has  
\eq
&&q_1^0\rightarrow (q_1\cdot v'),\quad
\vec q_1\rightarrow \vec q_1-q_1^0\vec v~'
+(v^{\prime 0}+1)^{-1}(\vec q_1\vec v~'\,)\vec v~'
\nonumber\\[2mm]
&&l_1^0\rightarrow (l_1\cdot v),\quad
\vec l_1\rightarrow \vec l_1-l_1^0\vec v
+({v}^{0}+1)^{-1}(\vec l_1\vec v\,)\vec v
\nonumber\\[2mm]
&&q_{23}^0\rightarrow (q_{23}\cdot v'),\quad
\vec q_{23}\rightarrow \vec q_{23}
-q_{23}^0\vec v~'+(v^{\prime 0}+1)^{-1}(\vec q_{23}\vec v~'\,)\vec v~'
\nonumber\\[2mm]
&&l_{23}^0\rightarrow (l_{23}\cdot v),\quad
\vec l_{23}\rightarrow \vec l_{23}
-l_{23}^0\vec v+({v}^{0}+1)^{-1}(\vec l_{23}\vec v\,)\vec v
\en
Under this transformation the wave functions of the final and initial baryon
are transformed to $C\bar\theta(v')\phi(\vec q_1\vec q_{23})$ and
$C\theta(v)\phi(\vec l_1\vec l_{23})$, whereas $\tilde g_0^{-1}(v')$
and $\tilde g_0^{-1}(v)$ transform to $M'_B-\sum\limits_{i=1}^3w'_i$
and $M_B-\sum\limits_{i=1}^3w_i$, respectively, with
\eq\label{momentum-prime}
&&w'_i=(m_i^2+{\vec p_i}^{~\prime 2})^{1/2},\quad\quad
  w_i=({m'_i}^{~2}+{\vec k_i}^{~\prime 2})^{1/2}
\nonumber\\[2mm]
&&{p_1'}=-\frac{1}{3}\, q_1,\quad
  {p_2'}=\frac{m_2}{3(m_2+m_3)}\, q_1-\frac{1}{2\sqrt{3}}\, q_{23},\quad
  {p_3'}=\frac{m_3}{3(m_2+m_3)}\, q_1+\frac{1}{2\sqrt{3}}\, q_{23}
\nonumber\\[2mm]
&&{k_1'}=-\frac{1}{3}\, l_1,\quad
  {k_2'}=\frac{m_2}{3(m_2+m_3)}\, l_1-\frac{1}{2\sqrt{3}}\, l_{23},\quad
  {k_3'}=\frac{m_3}{3(m_2+m_3)}\, l_1+\frac{1}{2\sqrt{3}}\, l_{23}
\en
After the change of integration variables one has
\eq\label{change_prop}
&&S^{(i')}(p_i)
\\[2mm]
&\to&\frac
{(\mu'_i M'_B+{{p_i'}^0})\not\! v'+w'_i-(m'_i+w'_i)^{-1}{\vec p_i}^{~\prime 2}
+({\vec p_i~'}\vec v~')(\gamma_0+(v^{\prime 0}+1)^{-1}(\vec\gamma\,\vec v~'))-
(\vec\gamma\,{\vec p_i~'}\,)}
{{w'_i}^2-(\mu'_i M'_B+{{p_i'}^0})^2-i0}
\nonumber\\[2mm]
&&S^{(i)}(k_i)\to
\\[2mm]
&\to&\frac
{(\mu_i M_B+{{k_i'}^0})\not\! v+w_i-(m_i+w_i)^{-1}{\vec k_i}^{~\prime 2}
+({\vec k_i~'}\vec v\,)(\gamma_0+({v}^{0}+1)^{-1}(\vec\gamma\,\vec v\,))-
(\vec\gamma\,{\vec k_i~'})}
{{w_i}^{~2}-(\mu_i M_B+{k_i}^{\prime 0})^2-i0}
\nonumber
\en
In the spectator approximation one neglects the transformed momenta 
${\vec p_i~'}$, ${\vec k_i'}$ in the numerators of Eq. (\ref{change_prop}).
Further, we can set $\not\! v\rightarrow 1$ and $\not\! v~'\rightarrow 1$
in the numerators since they act either on the final or initial spectator wave
functions. As a result one obtains in the spectator approximation
\eq
S^{(i')}(p_i)\rightarrow\frac{1}{w'_i-\mu'_i M'_B-{{p_i'}^0}-i0},\quad\quad
S^{(i)}(k_i)\rightarrow\frac{1}{w_i-\mu_i M_B-{{k_i'}^0}-i0}
\en
Further, in the heavy quark limit one neglects terms of order
$(2m_1)^{-1}{\vec p_1}^{~\prime 2}$ and
$(2m'_1)^{-1}{\vec k_1}^{~\prime 2}$. One then obtains the following effective
propagators
\eq
&&S^{(1')}(p_1)\rightarrow\frac{1}{-\bar\Lambda+m_2+m_3-{{p_1'}^0}-i0},
\quad\quad
S^{(1)}(k_1)\rightarrow\frac{1}{-\bar\Lambda+m_2+m_3-{{k_1'}^0}-i0}
\nonumber\\[2mm]
&&S^{(i)}(p_i)\rightarrow\frac{1}{w'_i-m_i-{{p_i'}^0}-i0},\quad\quad
S^{(i)}(k_i)\rightarrow\frac{1}{w_i-m_i-{{k_i'}^0}-i0},\quad\quad
i=2,3
\en
The factors $\tilde g_0^{-1}$ turn into
\eq
\tilde g_0^{-1}(v')\rightarrow\bar\Lambda-w'_2-w'_3,\quad\quad
\tilde g_0^{-1}(v)\rightarrow\bar\Lambda-w_2-w_3
\en
where all reference to the heavy quark mass has disappeared, as it should
indeed be. Further, substituting these expressions in Eq. (\ref{sought-for}),
one immediately arrives at the Eqs. (\ref{element}) and (\ref{fomega}).

It is interesting to note that one can write down "Feynman rules" for the
construction of the current matrix elements in the spectator approximation
which are remarkably simple since the vertices and propagators have a trivial
identity structure in the space of Lorentz indices. As an example we evaluate 
the current-induced transition between heavy baryons according to the diagram 
Fig.4 where we have added the appropriate vertex and propagator structure. The 
spectator model Feynman rules for this transitions may be summarized as 
follows: 
 
\noindent
1. A factor
$(\bar\Lambda -\sum\limits_{light}w_{light}(\vec p\,))
\phi(\vec p_1\vec p_2\vec p_3\,)$ for each heavy baryon vertex where 
$\phi(\vec p_1\vec p_2\vec p_3\,)$
denotes the radial part of the heavy baryon equal-time BS wave function

\noindent
2. The heavy quark propagator:
\eq
\frac{1}{-\bar\Lambda +\sum\limits_{light}m_{light}-{{p'}^0}-i0}
\en

\noindent
3. The light quark propagator:
\eq
\frac{1}{w'_i-m'_i-{{p_i'}^0}-i0}
\en

\noindent
4. Dirac $\delta$-functions corresponding to the four-momentum conservation
of light spectator quarks. All momenta are boosted to the rest frame of either
the final or the initial baryon.

\noindent
5. Integration over all relative four-momenta. 

\noindent
6. The spin-flavor structure of matrix elements is given by spectator model
wave function scalar products.

\newpage

\centerline{\bf TABLE CAPTIONS}

\noindent {\bf Table I.}
Heavy meson leptonic decay constants.
\vspace*{.2cm}

\noindent {\bf Table II.}
Slope of the heavy meson IW function.
\vspace*{.2cm}

\noindent {\bf Table III.}
Experimental and theoretical values for the branching
ratios (in \%) and asymmetry parameters in the decay
${B\rightarrow D(D^*)e\bar\nu}$.
\vspace*{.2cm}

\noindent {\bf Table IV.}
Exclusive decay rates of bottom baryons (in $10^{10}$ sec$^{-1}$)
for $|V_{bc}|=0.04$.
\vspace*{.2cm}

\noindent {\bf Table V.}
Asymmetry parameters for $\Lambda_b$ decay.
\vspace*{.2cm}

\vspace*{1.cm}

\centerline{\bf FIGURE CAPTIONS}

\noindent{\bf Fig. 1.}
Heavy baryon-heavy antibaryon pair production in the $e^+e^-$ annihilation:\\
a) Light quarks are produced independently from the vacuum by soft gluon
excahanges in the $J^P=0^+$ channel,\\
b) Tightly bound light diquark is produced from the vacuum.
\vspace*{.2cm}

\noindent{\bf Fig. 2.}
The heavy baryon IW function $\zeta(\omega)$ in the noninteracting light quark
model and in the full model for the different values of the wave function
range parameter $\Lambda_B$:\\
full model with $\Lambda_B=500$~MeV (solid line),\\
noninteracting light quark model with $\Lambda_B=500$~MeV (long-dashed line),\\
full model with $\Lambda_B=355$~MeV (short-dashed line).
\vspace*{.2cm}

\noindent{\bf Fig. 3.}
Relativistic effect in the heavy baryon IW function in full model with
$\Lambda_B=500$~MeV:\\
the function $\zeta(\omega)$; relativistic case (solid line),\\
the universal form factor function $f(\omega)$; relativistic case (long-dashed line),\\
the function $\zeta_{NR}(\omega)$; nonrelativistic case (short-dashed line).
\vspace*{.2cm}

\noindent{\bf Fig. 4.}
"Feynman rules" in the spectator quark approximation.

\newpage

\noindent
{\bf TABLE I}

\begin{center}
\begin{tabular}{| l l l l |}
 \hline
~Process            ~&~ Quantity    ~&~ Our ~&~ Lattice \cite{Lattice_1} ~\\
 \hline
~$D\to\ell\nu_\ell$ ~&~ $f_D$~(MeV) ~&~ 226 ~&~ $200\pm 30$           ~\\
~$B\to\ell\nu_\ell$ ~&~ $f_B$~(MeV) ~&~ 134 ~&~ $180\pm 40$           ~\\
 \hline
\end{tabular}
\end{center}

\noindent
{\bf TABLE II}

\begin{center}
\begin{tabular}{| l | l |}
 \hline
~$\rho^2$                    ~&~ Approach                                ~\\
 \hline
~1.00 (input)                ~&~ Our                                     ~\\
~$0.9^{+0.2+0.4}_{-0.3-0.2}$ ~&~ Lattice \cite{Baxter}                   ~\\
~$0.84\pm 0.02$              ~&~ QCD sum rules \cite{Bagan}              ~\\
~$0.70\pm 0.25$              ~&~ QCD sum rules \cite{Blok}               ~\\
~$0.42-0.92$                 ~&~ Quark Confinement Model \cite{Mizutani} ~\\
~$1.02$                      ~&~ Quasipotential \cite{Galkin}            ~\\
 \hline
\end{tabular}
\end{center}

\noindent
{\bf TABLE III}

\begin{center}
\begin{tabular}{| l | l | l |}
\hline
~                                           ~&~  Theory                 ~&~ Experiment                                        ~\\
 \hline
~$Br(B\rightarrow D)$                       ~&~  2.05 $|V_{bc}/0.04|^2$ ~&~ $1.6\pm 0.7$ \cite{PDG}, $1.9\pm 0.5$ \cite{PDG}  ~\\
 \hline
~$Br(B\rightarrow D^*)$                     ~&~  5.35 $|V_{bc}/0.04|^2$ ~&~ $6.6\pm 2.2$ \cite{PDG}, $4.4\pm 0.4$ \cite{PDG}  ~\\
 \hline
%~$Br(B\rightarrow D^*)/Br(B\rightarrow D)$  ~&~  2.61                   ~&~ $2.6^{+1.1+1.0}_{-0.6-0.8}$                      ~\\
% \hline
~$\alpha_{pol}$                             ~&~  1.71                   ~&~ $1.1\pm 0.4\pm 0.2$ \cite{ARGUS}                  ~\\
 \hline
~$\alpha'$                                  ~&~  0.63                   ~&~                                                   ~\\
 \hline
~$A_{FB}$                                   ~&~ 0.083                   ~&~ $0.20\pm 0.08\pm 0.06$ \cite{CLEOII}              ~\\
 \hline
~$A_{FB}^T$                                 ~&~  0.20                   ~&~                                                   ~\\
 \hline
\end{tabular}
\end{center}
\vspace*{.5cm}

\newpage
\noindent
{\bf TABLE IV}

\begin{center}
\begin{tabular}{|l|c|c|c|c|c|c|}
\hline
~Process                               ~&~ \cite{Singleton} ~&~ \cite{Cheng} ~&~ \cite{DESY} ~&~ \cite{Ivanov}     ~&~ Our (1) ~&~ Our (2) ~\\
  \hline
~$\Lambda_b^0\rightarrow\Lambda_c^+$   ~&~ 5.9              ~&~ 5.1          ~&~ 5.14            ~&~ 5.39          ~&~ 6.52    ~&~ 6.09    ~\\
~$\Xi_b^0\rightarrow\Xi_c^+$           ~&~ 7.2              ~&~ 5.3          ~&~ 5.21            ~&~ 5.27          ~&~ 6.83    ~&~ 6.42    ~\\
~$\Sigma_b^+\rightarrow\Sigma_c^{++}$  ~&~ 4.3              ~&~              ~&~                 ~&~ 2.23          ~&~ 1.90    ~&~ 1.65    ~\\
~$\Omega_b^-\rightarrow\Omega_c^0$     ~&~ 5.4              ~&~ 2.3          ~&~ 1.52            ~&~ 1.87          ~&~ 2.05    ~&~ 1.81    ~\\
~$\Sigma_b^+\rightarrow\Sigma_c^{*++}$ ~&~                  ~&~              ~&~                 ~&~ 4.56          ~&~ 4.17    ~&~ 3.75    ~\\
~$\Omega_b^-\rightarrow\Omega_c^{*0}$  ~&~                  ~&~              ~&~ 3.41            ~&~ 4.01          ~&~ 4.55    ~&~ 4.13    ~\\
  \hline
\end{tabular}
\end{center}
\vspace*{.5cm}

\noindent
{\bf TABLE V}

\begin{center}
\begin{tabular}{|l|c|c|c|c|c|c|}
 \hline
~              ~&~ $\alpha$ ~&~ $\alpha'$ ~&~ $\alpha''$ ~&~ $\gamma$ ~&~ $\alpha_P$ ~&~ $\gamma_P$ ~\\
 \hline
~Our (1)       ~&~ -0.78    ~&~ -0.11     ~&~ -0.55      ~&~ 0.54     ~&~ 0.41       ~&~ -0.15      ~\\
~Our (2)       ~&~ -0.78    ~&~ -0.11     ~&~ -0.55      ~&~ 0.54     ~&~ 0.41       ~&~ -0.16      ~\\
~\cite{Ivanov} ~&~ -0.76    ~&~ -0.12     ~&~ -0.53      ~&~ 0.56     ~&~ 0.39       ~&~ -0.17      ~\\
~\cite{Konig}  ~&~ -0.74    ~&~ -0.12     ~&~ -0.46      ~&~ 0.61     ~&~ 0.33       ~&~ -0.19      ~\\
 \hline
\end{tabular}
\end{center}

\newpage
\begin{figure}
\end{figure}

\vspace*{2.cm}

\unitlength=1mm
\special{em:linewidth 0.4pt}
\linethickness{0.4pt}
\begin{picture}(144.00,81.00)
\put(35.00,35.00){\vector(-1,1){15.00}}
\put(35.00,35.00){\vector(1,1){15.00}}
\put(35.00,45.00){\vector(-1,1){15.00}}
\put(35.00,45.00){\vector(1,1){15.00}}
\put(35.00,55.00){\vector(-1,1){15.00}}
\put(35.00,55.00){\vector(1,1){15.00}}
\put(20.00,50.00){\line(-1,1){10.00}}
\put(20.00,60.00){\line(-1,1){10.00}}
\put(20.00,70.00){\line(-1,1){10.00}}
\put(50.00,50.00){\line(1,1){10.00}}
\put(50.00,60.00){\line(1,1){10.00}}
\put(50.00,70.00){\line(1,1){10.00}}
\put(35.00,35.00){\line(0,-1){3.00}}
\put(35.00,30.00){\line(0,-1){3.00}}
\put(35.00,25.00){\line(0,-1){3.00}}
\put(40.00,40.00){\circle*{2.00}}
\put(40.00,43.00){\circle*{2.00}}
\put(40.00,46.00){\circle*{2.00}}
\put(40.00,49.00){\circle*{2.00}}
\put(30.00,40.00){\circle*{2.00}}
\put(30.00,43.00){\circle*{2.00}}
\put(30.00,46.00){\circle*{2.00}}
\put(30.00,49.00){\circle*{2.00}}
\put(30.00,52.00){\circle*{2.00}}
\put(30.00,55.00){\circle*{2.00}}
\put(30.00,58.00){\circle*{2.00}}
\put(44.00,48.00){\makebox(0,0)[cc]{$0^+$}}
\put(26.00,59.00){\makebox(0,0)[cc]{$0^+$}}
\put(6.00,60.00){\makebox(0,0)[cc]{$\frac{1}{2}^+$}}
\put(6.00,70.00){\makebox(0,0)[cc]{$\frac{1}{2}^+$}}
\put(6.00,80.00){\makebox(0,0)[cc]{$\frac{1}{2}^+$}}
\put(64.00,60.00){\makebox(0,0)[cc]{$\overline{\frac{1}{2}^+}$}}
\put(64.00,70.00){\makebox(0,0)[cc]{$\overline{\frac{1}{2}^+}$}}
\put(64.00,80.00){\makebox(0,0)[cc]{$\overline{\frac{1}{2}^+}$}}
\put(115.00,35.00){\vector(1,1){15.00}}
\put(115.00,35.00){\vector(-1,1){15.00}}
\put(115.00,35.00){\line(0,-1){3.00}}
\put(115.00,30.00){\line(0,-1){3.00}}
\put(115.00,25.00){\line(0,-1){3.00}}
\put(130.00,50.00){\line(1,1){10.00}}
\put(100.00,50.00){\line(-1,1){10.00}}
\put(115.00,54.00){\line(1,1){25.00}}
\put(115.00,54.00){\line(-1,1){25.00}}
\put(115.00,56.00){\line(1,1){25.00}}
\put(115.00,56.00){\line(-1,1){25.00}}
\put(120.00,40.00){\circle*{2.00}}
\put(120.00,43.00){\circle*{2.00}}
\put(120.00,46.00){\circle*{2.00}}
\put(120.00,49.00){\circle*{2.00}}
\put(120.00,52.00){\circle*{2.00}}
\put(120.00,55.00){\circle*{2.00}}
\put(120.00,58.00){\circle*{2.00}}
\put(125.00,51.00){\makebox(0,0)[cc]{$0^+$}}
\put(85.00,80.00){\makebox(0,0)[cc]{$0^+$}}
%\put(64.00,60.00){\makebox(0,0)[cc]{$\frac{1}{2}^+$}}
\put(144.00,60.00){\makebox(0,0)[cc]{$\overline{\frac{1}{2}^+}$}}
\put(85.00,60.00){\makebox(0,0)[cc]{$\frac{1}{2}^+$}}
\put(144.00,80.00){\makebox(0,0)[cc]{$\overline{0^+}$}}
\put(35.00,10.00){\makebox(0,0)[cc]{{\bf a}}}
\put(115.00,10.00){\makebox(0,0)[cc]{{\bf b}}}
\end{picture}

\vspace*{3.cm}
\centerline{\bf Fig. 1}

\newpage
\begin{figure}
\begin{center}
%\vspace*{-70mm}
{\bf
\mbox{\epsfysize=18cm\epsffile{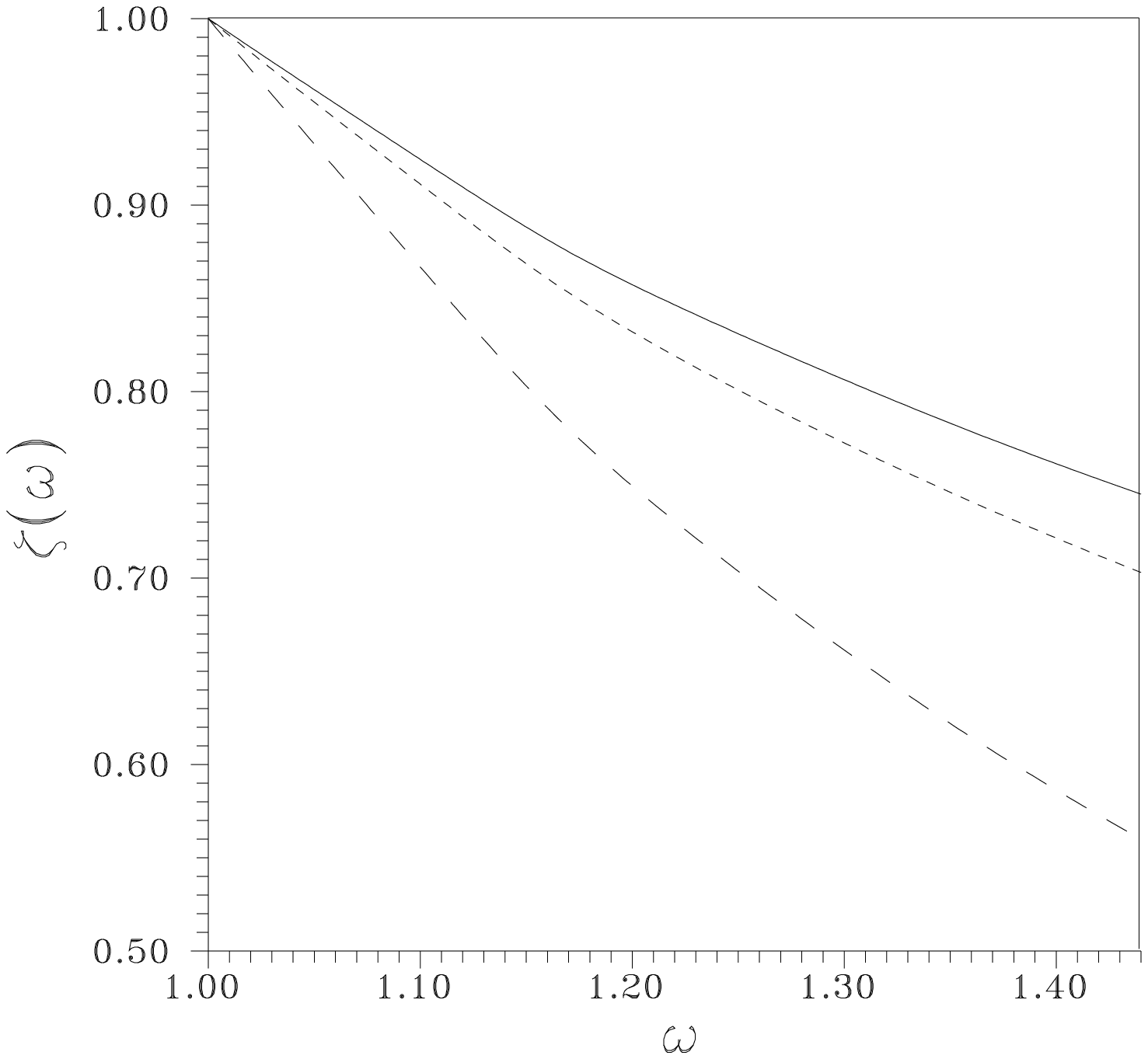}}}
\\
\vspace*{1.cm}
\mbox{\hspace*{-0.3cm}{\bf Fig. 2}}
\\
\end{center}
\end{figure}

\newpage
\begin{figure}
\begin{center}
%\vspace*{-70mm}
{\bf
\mbox{\epsfysize=18cm\epsffile{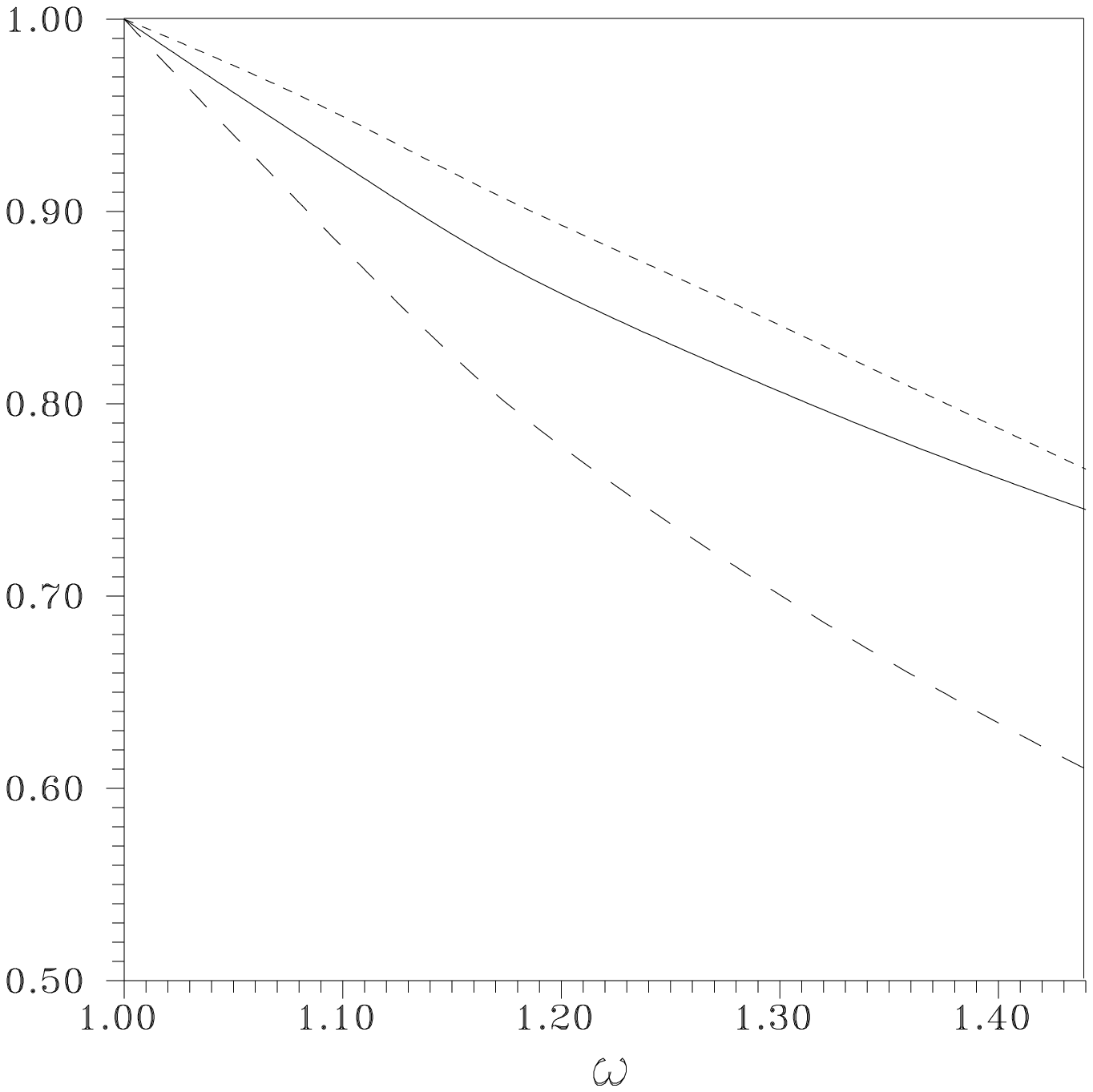}}}
\\
\vspace*{1.cm}
\mbox{\hspace*{-0.3cm}{\bf Fig. 3}}
\\
\end{center}
\end{figure}

\newpage

\unitlength=1mm
\special{em:linewidth 0.4pt}
\linethickness{0.4pt}
\hspace*{1.cm}\begin{picture}(30.00,90.00)
\put(47.50,50.00){\oval(20.00,20.00)[l]}
\put(47.50,60.00){\line(0,-1){20.00}}
\put(7.50,50.00){\line(1,0){30.00}}
\put(7.50,51.00){\line(1,0){30.04}}
\put(7.50,49.00){\line(1,0){30.04}}
\put(20.50,50.00){\line(-1,1){4.00}}
\put(20.50,50.00){\line(-1,-1){4.00}}
\put(47.50,42.00){\vector(1,0){20.00}}
\put(67.50,42.00){\line(1,0){20.00}}
\put(47.50,50.00){\vector(1,0){20.00}}
\put(67.50,50.00){\line(1,0){20.00}}
\put(87.50,50.00){\oval(20.00,20.00)[r]}
\put(87.50,60.00){\line(0,-1){20.00}}
\put(127.50,50.00){\line(-1,0){30.00}}
\put(127.50,51.00){\line(-1,0){30.04}}
\put(127.50,49.00){\line(-1,0){30.04}}
\put(117.50,50.00){\line(-1,1){4.00}}
\put(117.50,50.00){\line(-1,-1){4.00}}
\put(47.50,58.00){\vector(1,1){11.00}}
\put(67.50,78.00){\line(-1,-1){9.00}}
\put(67.50,78.00){\vector(1,-1){11.00}}
\put(87.50,58.00){\line(-1,1){9.00}}
\put(67.50,78.00){\circle*{2.00}}
\put(67.50,78.00){\line(0,1){1.50}}
\put(67.50,80.00){\line(0,1){1.50}}
\put(67.50,82.00){\line(0,1){1.50}}
\put(67.50,84.00){\line(0,1){1.50}}
\put(2.00,50.00){\makebox(0,0)[cc]{${\bf P}$}}
\put(131.00,50.00){\makebox(0,0)[cc]{${\bf P'}$}}
\put(67.50,35.00){\makebox(0,0)[cc]{$\frac{1}{w'_3-m_3-{{p_3'}^0}-i0}$}}
\put(67.50,55.00){\makebox(0,0)[cc]{$\frac{1}{w'_2-m_2-{{p_2'}^0}-i0}$}}
\put(92.50,70.00){\makebox(0,0)[cc]{$\frac{1}{-\bar\Lambda+m_2+m_3-{{p_1'}^0}-i0}$}}
\put(42.50,70.00){\makebox(0,0)[cc]{$\frac{1}{-\bar\Lambda+m_2+m_3-{{k_1'}^0}-i0}$}}
\put(107.50,28.00){\makebox(0,0)[cc]{$(\bar\Lambda-w'_2-w'_3)\phi(\vec q_1\vec q_{23})$}}
\put(28.50,28.00){\makebox(0,0)[cc]{$\phi(\vec l_1\vec l_{23})(\bar\Lambda-w_2-w_3)                         $}}
\put(97.50,40.00){\vector(1,-1){8.00}}
\put(37.50,40.00){\vector(-1,-1){8.00}}
\put(67.50,-5.00){\makebox(0,0)[cc]{{\bf Fig. 4}}}
\end{picture}

\end{document}